\documentclass{sig-alternate-10pt}
\usepackage{amsmath,epsfig}
\usepackage{subfigure}
\usepackage[hyphens]{url}
\usepackage{verbatim}
\usepackage{array}
\title{The Pin-Bang Theory: Discovering The Pinterest World}

\author{%
{Sudip Mittal, Neha Gupta
, Prateek Dewan, Ponnurangam Kumaraguru }%
\vspace{1.6mm}\\
Indraprastha Institute of Information Technology, Delhi (IIIT-D)\\
\{sudip09068, neha1209, prateekd, pk\}@iiitd.ac.in%
\vspace{1.2mm}\\
}

\begin{document}
\maketitle

\begin{abstract}

Pinterest is an image-based online social network, which was launched in the year 2010 and has gained a lot of traction, ever since. Within 3 years, Pinterest has attained 48.7 million unique users. This stupendous growth makes it interesting to study Pinterest, and gives rise to multiple questions about it's users, and content. We characterized Pinterest on the basis of large scale crawls of 3.3 million user profiles, and 58.8 million pins. In particular, we explored various attributes of users, pins, boards, pin sources, and user locations, in detail and performed topical analysis of user generated textual content. The characterization revealed most prominent topics among users and pins, top image sources, and geographical distribution of users on Pinterest. We then investigated this social network from a privacy and security standpoint, and found traces of malware in the form of pin sources. Instances of \emph{Personally Identifiable Information} (PII) leakage were also discovered in the form of phone numbers, ~BBM ~(Blackberry Messenger) pins, and~ email addresses. Further, our analysis demonstrated how Pinterest is a potential venue for copyright infringement, by showing that almost half of the images shared on Pinterest go uncredited. 
To the best of our knowledge, this is the first attempt to characterize Pinterest at such a large scale.

\end{abstract}



\section{Introduction}
%

Online Social Networks (OSNs) like Facebook, Twitter, LinkedIn, and Google+ are web-based platforms that help users to interact, share thoughts, interests, and activities. These OSNs allow their users to imitate real life connections over the Internet. A report by the International Telecommunication Union states that the total number of online social media users has crossed the 1 billion mark as of May, 2012~\cite{Lunden:2012a}. According to Nielsen's Social Media Report, users continue to spend more time on social networks than on any other kind of websites on the Internet~\cite{Nielsen:2012}. With this outburst in the number of social media users across the world, online social media has moved to the next level of innovation. While all the aforementioned conventional social media services are mostly text-intensive, some of them have gone beyond text, and have introduced images as their building blocks. Services like Instagram, and Tumblr have gained immense popularity in the recent years, with Instagram (now part of Facebook) attaining 100 million monthly active users, and 40 million photo uploads per day~\cite{Center:2013}. 
These numbers indicate successful entrance of image based social networks in the world of online social media.

Pinterest is one of the most recent additions to this popular category of image-based online social networks. Within a year of its launch, Pinterest was listed among the ``50 Best Websites of 2011'' by Time Magazine~\cite{McCracken:2011}. It was also the fastest site to break the 10 million unique visitors mark~\cite{Constine:2012}. Number of users since then have increased, with Reuters stating a figure of 48.7 million unique users in February 2013~\cite{Reuters:2013}. Although fairly new to the social media fraternity, Pinterest is being heavily used by many big business houses like Etsy, The Gap, Allrecipes, Jettsetter, etc. to advertise their products.~\footnote{\url{http://business.pinterest.com/stories/}}
Further, Pinterest drives more revenue per click than Twitter or Facebook, and is currently valued at USD 2.5 billion~\cite{Reuters:2013,Zwelling:2012}.

The immense upsurge and popularity of Pinterest has given rise to multiple basic questions about this network. 
What is the general user behavior on Pinterest? What are the most common characteristics of users, pins, and boards? What is the sentiment associated with user-generated textual content? What is the geographical distribution of users? How well is Pinterest safeguarded against privacy and security threats?
There exists little research work on Pinterest~\cite{Ana-Maria-Popescu:2013,carpenter2012copyright,kamath2013board,ottoni2013ladies,zarro2012pinterest,zarro2013wedding}; but none of this work addresses the aforementioned basic questions. To answer these questions, and get deeper insights into Pinterest, we collected and analyzed a dataset comprising of user details (3,323,054), pin details (58,896,156), board details (777,748), and images (498,433). 

To the best of our knowledge, this is one of the first large-scale studies conducted on Pinterest, exploring it's users, and content. 
Based on our analysis, some of our key contributions are summarized as follows:
\begin{enumerate}
\item Topical analysis of user generated textual content on Pinterest: We found that the most common topics across users, and pins were design, fashion, photography, food, and travel.
\item User, pin, and board characterization: We analyzed various user profile attributes, their geographical distribution, top pin sources, and board categories. Less than 5\% of all images on Pinterest are uploaded by users; over 95\% are pinned from pre-existing web sources.

\item Exploring Pinterest as a possible venue for copyright infringement: We found copyrighted images being shared publicly on Pinterest, and almost half of these images did not give due credit to the copyright owners.

\item Analysis of personal information and malicious content present on Pinterest: Users were found to leak a significant amount of \emph{Personally Identifiable Information (PII)} voluntarily. We found numerous instances where users shared their phone numbers, Blackberry Messenger (BBM) pins, email IDs, marital status, and other personal information, publicly. We also found (and analyzed) traces of malware in the form of pin sources.

\end{enumerate}
The rest of the paper is organized as follows. We discuss the related work in Section~\ref{section:pinterest:relatedwork}. We then discuss Pinterest as a social network in Section~\ref{section:pinterestabout}. In Section~\ref{section:methodology}, we describe our data collection methodology. Analysis of the collected data and its results are covered in Section~\ref{section:analysis}. In Section~\ref{section:privacy_security}, we explore various privacy and security implications on Pinterest. 
Section~\ref{section:discussion} contains discussion, limitations, and future work.
\section{Related Work}\label{section:pinterest:relatedwork}


Online social networks, in general, have been studied in detail by various researchers in the computer science community. Mislove et al. conducted a large scale measurement study and analysis of Flickr, YouTube, LiveJournal, and Orkut~\cite{mislove2007measurement}. Their results confirmed power-law, small-world, and scale-free properties of online social networks. In a more recent work by Magno et al., authors performed a detailed analysis of the Google+ network, and identified some key differences and similarities between Google+, and existing social networks like Facebook, and Twitter~\cite{magno2012new}. Ugander et al. performed a large-scale analysis of the entire Facebook social graph and found
that 99.91\% of all the users belonged to a single large connected component~\cite{ugander2011anatomy}. They confirmed the `six degrees of separation' phenomenon
and showed that the value had dropped to 3.74 degrees of separation in the entire Facebook network of active users.

Although there has been a lot of work done on the network structure, and user characteristics on various social networks, to the best of our knowledge, little work has been done on image-based social networks; in particular, Pinterest. Hochman et al.~\cite{hochman2012visualizing} used Cultural Analytics visualization techniques to study 550,000 images taken by users of the social photo sharing application, Instagram. Authors of this work compared the images from New York City and Tokyo, and found differences in local color usage, cultural production rate, etc. Hollenstein et al. harvested geo-referenced and tagged metadata associated with 8 million Flickr images, and considered how large numbers of people named city core areas~\cite{hollenstein2013exploring}. Authors of this work exploited the fact that the terms used to describe city centers, such as Downtown, are key concepts in everyday or vernacular language. The study covered six cities around the world, viz. Zurich, London, Sheffield, Chicago, Seattle, and Sydney.

Considering the rapid growth rate of Pinterest since its launch, there still exist only a few studies on this social network. Ottoni et al. analyzed Pinterest in a gender-sensitive fashion, and found that the network was heavily dominated by female users. Authors of this work found that females on Pinterest make more use of lightweight interactions than males, invest more effort in reciprocating social links, are more active and generalist in content generation, and describe themselves using words of affection and positive emotions. This study spanned across a large dataset consisting of over 2 million users~\cite{ottoni2013ladies}. Kamath et al. described a supervised model for board recommendation on Pinterest. They used a content-based filtering approach for recommending high quality information to users~\cite{Krishna-Kamath:2013}. Dudenhoffer et al. tried to use Pinterest as a library marketing and information literacy tool at the Central Methodist University. They reported that the number of followers viewing the library pinboards had outpaced usage of the text-based lists in just one semester~\cite{dudenhoffer2012pin}. In another similar work by Zarro et al., authors talked about how digital libraries and other organizations could take advantage of Pinterest to expand the reach of their material, allowing users to create personalized collections, incorporating their content~\cite{zarro2012pinterest}. Carpenter evaluated the potential copyright infringement liability for Pinterest users, and stated that the statutory fair use defense~\footnote{\url{http://www.copyright.gov/fls/fl102.html}} is likely to shield most Pinterest users from copyright infringement liability~\cite{carpenter2012copyright}. 

None of the aforementioned work on Pinterest concentrated on characterizing Pinterest as a social network, but only used Pinterest 
to study other related problems. In this work, we intend to characterize Pinterest, and get insights on the users who use it, their interests, ~the ~content ~they ~generate, ~major content sources, top user locations, and security and privacy issues on Pinterest. 

\section{Understanding Pinterest}\label{section:pinterestabout}
Pinterest is an image-based social bookmarking media, where users share images which are of interest to them, in the form of~\emph{pins} on a~\emph{pinboard}. It emphasizes on~\emph{discovery} and~\emph{curation} of images rather than original content creation.~\footnote{\url{http://blogs.constantcontact.com/product-blogs/social-media-marketing/what-the-heck-is-pinterest-and-why-should-you-care/}} 
This makes Pinterest a very promising conduit for the promotion of commercial activities online. 

Similar to other OSNs,
Pinterest also uses some specific terminology to refer to various elements and services it provides. Some terms are as follows:

\begin{enumerate}
\item {\bf Pins}: A~\emph{pin} is an image that has some meta-data information associated with it. Pins can be thought of as basic building blocks of Pinterest. The act of posting a~\emph{pin} is known as~\emph{pinning}, and the user who posts a pin is the~\emph{pinner}. Similar to images on Facebook, pins can be~\emph{liked} and~\emph{shared}. Each of these pins has the following meta-data associated
with it -- \emph{unique pin number}, \emph{description}, \emph{number of likes}, \emph{number of comments}, \emph{number of repins}, \emph{board name}, \emph{source}, and \emph{content in comments}. The act of sharing an already existing pin is referred to as~\emph{repinning}.
\item {\bf Pinboards}: They are a themed collection of pins, organized by a user. Each board (``boards" and ``pinboards" are used interchangeably) has a name, a description (optional), category (optional, e.g. Animals, Art, Celebrities, Food and Drink, Design, ~Education, ~Gardening), and an option to make it Secret. \emph{Secret boards} are only visible to the users who create them. 
This analogy of pins and pin-boards replicates the real-world concept of images on a scrapbook.
\item {\bf Source}: Each~\emph{pin} on Pinterest has a source URL associated with it. As the name suggests, this is the actual URL from which the image has been pinned by a user. Images uploaded by users directly to Pinterest from their local computer, have ~\emph{pinterest.com} as ~their ~source, whereas images which are pinned from an existing website (e.g. flickr.com) have this source website (flickr.com) as the source.
\item {\bf Pin-It button}: A~\emph{Pin-It} button is a browser bookmark used to upload content to Pinterest. Some popular websites like Amazon, eBay, BHG, and Etsy also provide their own pin-it button next to their product images. This pin-it button makes it easier for a user to share the content that she likes on Pinterest. 
\end{enumerate}
\subsection{User Accounts}
\begin{figure}[!ht]
\centering
\fbox{\includegraphics[scale=0.5]{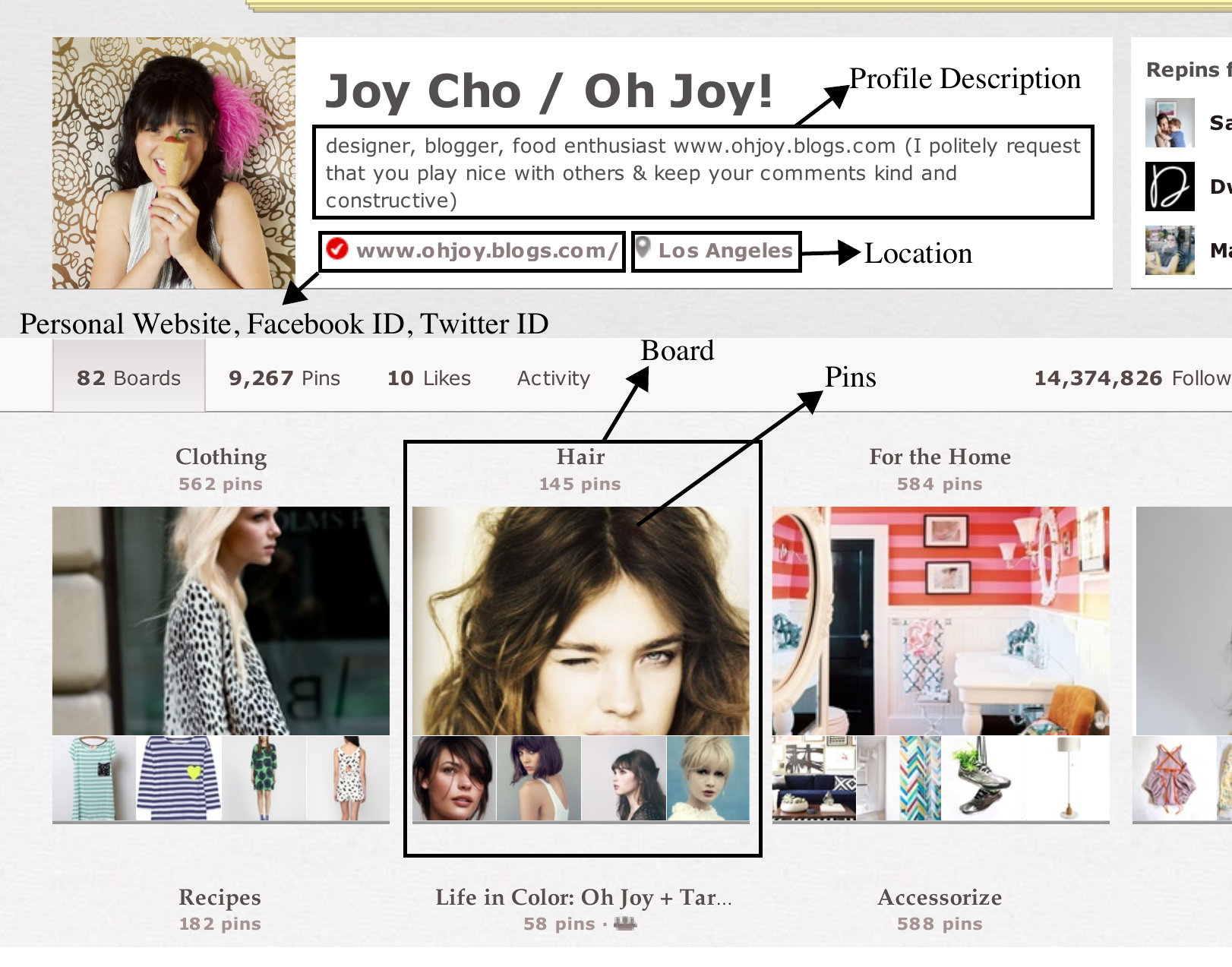}}
\caption{User Profile on Pinterest. The \emph{profile description, websites, location, board, and pin} are marked separately in the screen snapshot.}
\label{fig:profile}
\end{figure}

A user begins by creating an account using her Facebook ID, Twitter ID, or an email address. On account creation, Pinterest asks each new user to follow 5 boards to complete the creation process, as a mandatory step to get started. 
Each user has a profile page (Figure~\ref{fig:profile}) that is publicly visible to everyone, listing the user's name, a description, location, connected Facebook account (if available)
, connected Twitter account (if available), a profile website, boards (which are not secret), and associated pins, likes, followers, and followees. A user also has a timeline where all pins from the users she follows, are displayed.

%

\subsection{Social Ties}
A user has the option to follow a particular user or a specific board of any other user. If a user follows another user, she gets updates about all the boards owned by that user. But, in case a user follows specific boards, she gets updates only from those particular boards. This relationship is quite similar to Twitter's follower / followee relationship.
Interactions on Pinterest are in the form of pins. A user pins an image, and can add a pin description to better describe the pin. Other users can then repin the shared pin, like it or share their views through a comment. These features are similar to Facebook's share, like, and comment features respectively.

\subsection{Comparison with other OSNs}
Pinterest has now gained the reputation of being one of the most popular social media websites. With this immense popularity, it is quite often compared to the other leading OSNs like Facebook, Twitter, Instagram, etc. Thus, it is important to understand the basic similarities and differences between Pinterest and other such leading social platforms.
\subsubsection{Facebook and Twitter}
Unlike most other OSNs, Pinterest is an extensive photo sharing platform. Services like Facebook and Twitter follow a text dominant approach; Pinterest follows an image dominant approach. The amount of personal user information that Pinterest holds, is also less as compared to other networks. This can be clearly understood from the fact that Pinterest does not even ask it's users for basic information like gender, date of birth, phone number, etc. while creating the account. However, users are free to connect their profiles on other OSNs like Facebook and Twitter with their Pinterest profile. 
With all the boards and pins public (other than secret boards), Pinterest does not restrict its users to like, comment, or re-pin any of the available pins. This is quite similar to Twitter, but Facebook follows a different methodology. Since there are many more options on Facebook to ensure user privacy, a user has more control to what others see on his profile. Pinterest, on the other hand provides no such privacy options. 
\subsubsection{Instagram}
The basic difference between ~Pinterest and Instagram~\footnote{\url{http://instagram.com}} is that while Instagram is majorly for content creators, Pinterest is for curators. i.e people use Instagram to deliver the content, but Pinterest to share the content with public thereafter.~\footnote{\url{http://thewebadvisors.ca/visual-throwdown-instagram-v-pinterest/}} Also, a user can apply digital filters on the uploaded images on Instagram, but no such feature is available on Pinterest, yet. Pinterest provides a strong commercial prospects by directly linking a pin to the commercial website where the product presented on the pin can be purchased. This accounts for a stronger e-commerce behavior on Pinterest, which Instagram does not provide.
\subsubsection{Tumblr}
In contrast to Pinterest, Tumblr~\ is a microblogging application. On tumblr, one has a flexibility to post a short text, audio, image, and video, but Pinterest is only restricted to images and videos. Where on one hand Tumblr provides a single column display of images, Pinterest has a 5 column display, thus a user can view more images on a single page on Pinterest.~\footnote{\url{http://amandaaaa02.wordpress.com/2012/11/13/pinterest-social-media-sites/}} Tumblr and Pinterest also differ in terms of the target audience. Pinterest is used heavily by women in the age group of 25 and 44, whereas Tumblr is targeted to younger people in the age group of 18 and 25.~\footnote{\url{http://sproutsocial.com/insights/2012/02/pinterest-vs-tumblr/}}
\subsubsection{Reddit}
Reddit is a social news sharing and entertainment website where each user contributing the content is called a~\emph{redditor}, similar to a~\emph{pinner} on Pinterest. Such content is then socially curated on Reddit, the way it is done on Pinterest. Both these OSNs also follow a similar kind of virtual bulletin board like structure. Difference comes in terms of the target audience. Where on one hand Pinterest is a female dominant network, Reddit is a completely male dominant network.~\footnote{\url{http://www.webpronews.com/men-are-from-reddit-women-are-from-pinterest-infographic-2012-07}} Also, in terms of sharing views for an individual content, Reddit provides a~\emph{upvote/downvote} option that clearly represents a positive or negative public behavior towards the posted content. Pinterest on the other hand only provides a like option that can only capture a positive inclination of people towards the pin.

\section{Data collection}\label{section:methodology}

In this section, we discuss the methodology that we applied for data collection, and describe the data that we collected. Given the size of the entire Pinterest network (48.7 million users), it would have been hard, and computationally very expensive to be able to capture the entire network. 

\begin{figure}[!ht]
\centering
\fbox{\includegraphics[scale=0.34]{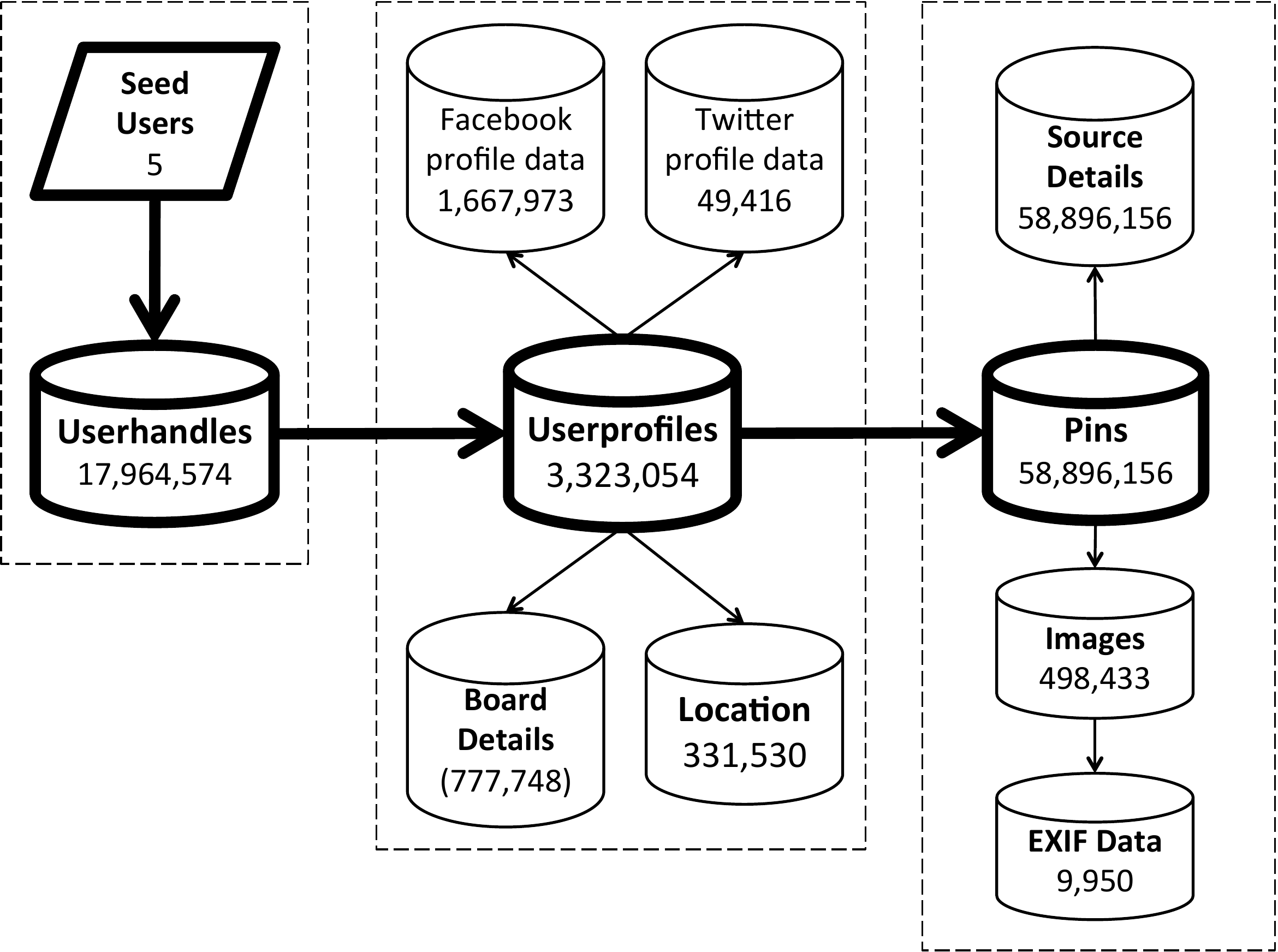}}
\caption{Flow diagram depicting the flow sequence of our data collection process. The darkened blocks represent our initial seed users, and primary datasets. The lighter blocks denote the additional information extracted through the primary dataset.}
\label{fig:crawler}
\end{figure}

Pinterest does not provide a public API for data collection. Therefore, in order to collect data, we designed and implemented a breadth first search (BFS) crawler in Python. All data was collected using a Dell PowerEdge R620 server, with 64 Gigabytes of RAM, 24 core processor, connected to a 1 Gbps Internet connection. The entire data collection process spanned from December 26, 2012 to February 1, 2013. 
Broadly, this process (Figure~\ref{fig:crawler}) was split into three phases as described below:
\subsection{User Handles Collection}
The data collection process was initiated by selecting the top 5 profiles in terms of the number of followers on Pinterest, as initial seeds, and feeding them into the crawler.
The crawler first extracted 4,995,974 direct followers of these 5 input seeds, and then repeatedly crawled through the ``followers of followers". We collected a total of 17,964,574 unique user handles through this process, which is slightly over 36\% of the entire Pinterest population~\cite{Reuters:2013}. We call this, the \emph{userhandles} dataset. 
This technique of snowball-sampling is commonly used in online social media research~\cite{ottoni2013ladies}. 

\subsection{User Data Collection}

Next, we started data collection for user profiles of the 17.96 million user handles collected in the previous step, 
and obtained a total of 3,323,054 user profiles, called the \emph{userprofile dataset} (we present the analysis on 3.3 million userprofiles in this paper; though our data collection process is still active). This userprofile dataset includes user display name, description field, profile picture, number of followers, number of followees, number of boards, number of pins, boards, profile website, Facebook handle, Twitter handle, location, pins, and likes. Along with user profiles, we extracted 777,748 boards and their corresponding details (called the \emph{boards dataset}). These details include board category, number of followers, and number of pins for each pinboard.

Many times users also mention their Facebook and / or Twitter profile URLs on Pinterest. Using this information from the userprofile dataset
, we collected publicly available Facebook information of 1,667,973 users (50.19\% of the userprofile dataset) and Twitter information of 49,416 users (1.4\% of the userprofile dataset). Many Pinterest users also mention location in their profile. We found location details for 331,530 users (9.93\% of the userprofile dataset). Some users mentioned only their country, whereas others mentioned their city as well. Some users gave their location as ``The beach", ``mentally in lala land", etc. In order to verify the credibility of such location information, we used \emph{Yahoo Placefinder API}~\footnote{\url{http://developer.yahoo.com/boss/geo/docs/requests-pf.html}} and obtained the correct details for 192,261 (57.99\%) of these locations.

\subsection{Pin Data Collection}
Using user profiles as seeds, we collected 58,896,156 unique pins and their related information. We call this the \emph{pin dataset}. This information consists of the pin description, number of likes, number of comments, number of repins, board name, and source for each pin. We also collected a random sample of 498,433 images (called the \emph{images dataset}) from these pins. 
For each of these images, we extracted their \emph{Exchangeable Image File Format} (EXIF) information for further analysis.~\footnote{\url{http://fotoforensics.com/tutorial-meta.php#EXIF}} Most common pieces of EXIF information available were date, time, image description, artist, copyright, and camera make / model. We also extracted information about pin sources for each pin, referred to as the \emph{source dataset}.
\section{Analysis}\label{section:analysis}

We now present our analysis of the users, pins, and boards in detail.

\subsection{User characterization}\label{sec:user}


\subsubsection{Profile description}\label{subsec:profiledescription}

\begin{figure}[!h]
\centering
\includegraphics[scale=0.36]{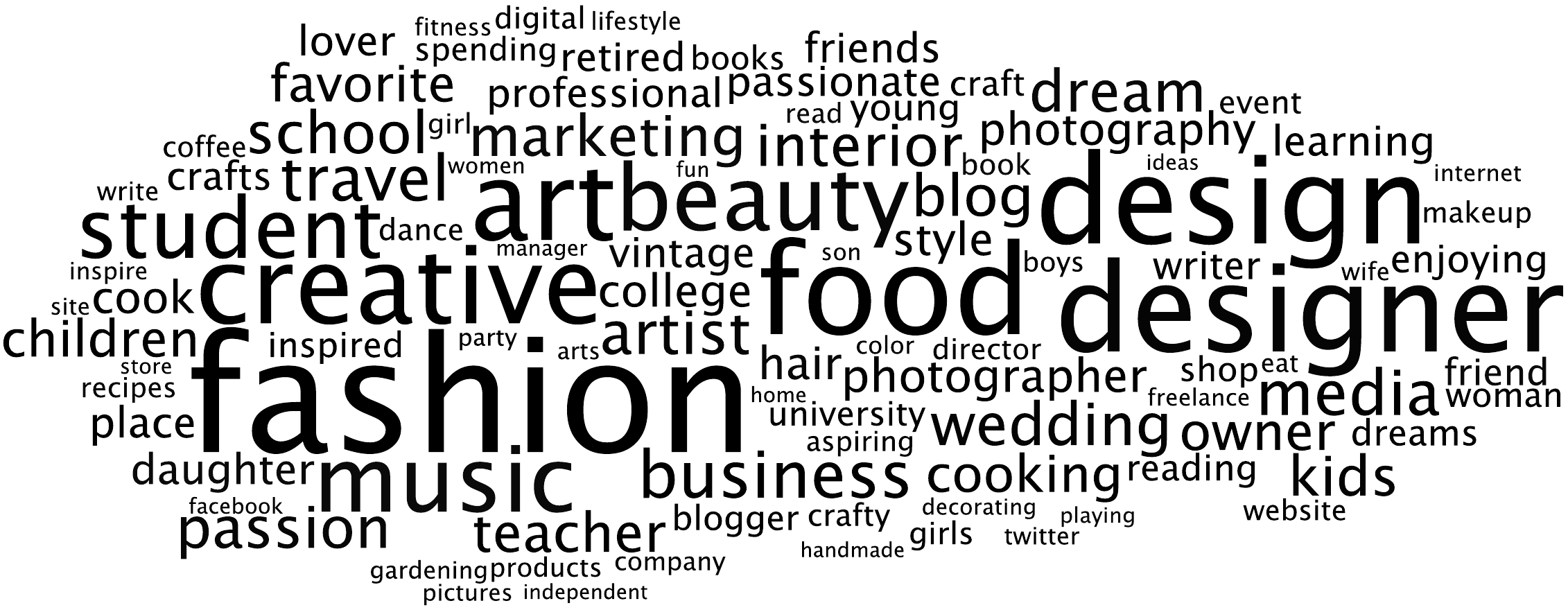}
\caption{Tag cloud of the top 100 words taken from user's profile description field.}
\label{fig:About}
\end{figure}

From our userprofile dataset of 3,323,054 user profiles, we found that only 589,193 (17.73\%) users had profile description. We observed that users revealed private details through this field, like age, marital status, personal traits, email IDs, phone numbers, etc.
The profile description of one user said, \emph{``I am 35, happily married, love kids \& cats, and have a disturbing sense of humor!"} We extracted 100 most frequently occurring words from the profile description, and found topics like fashion, design, food, music, art, photography, and travel as the most popular user interests (Figure~\ref{fig:About}). We observed that the most common interests 
were in line with the most common professions (like artist, designer, cook, photographer) mentioned by the users. This shows that large proportion of Pinterest consumers make use of the network for professional activities.



\subsubsection{Social and commercial links}
Another source of information on the user profile is the ``website" field, where users can provide URLs to their personal websites, and blogs. In our dataset, we found that 177,462 (5.34\%) users had mentioned a website. The topmost domain was Facebook, where 9,697 (5.46\%) users had mentioned a link to their Facebook profiles. Twitter, Etsy, YouTube, Flickr, About.me, LinkedIn, etc. were the other domains which constitute the top 10. 
Apart from the website field, Pinterest separately provides users with an option to connect their Facebook and / or Twitter accounts with their Pinterest profiles. Out of over 3.3 million user profiles that we collected, over 2.71 million users (81.78\%) had connected their Facebook profiles with Pinterest. Only 328,570 (9.88\%) users connected their Twitter accounts with Pinterest.
Less than 4\% (132,553) users had connected both Facebook and Twitter, while 12.3\% (409,399) users had connected neither. Further, we found that 86,641 (26.36\%) out of 328,570 users had identical usernames on Twitter and Pinterest, and 33,412 (10.17\%) had a similarity of more than 90\%. However, 5,419 (5.02\%) out of 107,910 users had identical usernames on Facebook and Pinterest, and only 1,253 (1.16\%) of these users had a similarity of more than 90\%. Two hundred and ninety seven (0.22\%) users had identical usernames on all three networks. Analysis of usernames for the same user on various social networks can be useful for identity resolution across multiple OSNs~\cite{jain2013seek}.


\subsubsection{Connections and popularity}
The maximum number of followers for a user was found to be 11,992,745 (as of January 2013). Table~\ref{tb:top_profiles} lists the description, number of followers and followees for the top 10 most followed users on Pinterest (to maintain users' privacy, we do not mention usernames anywhere).  
The average number of followers per Pinterest user was found to be approximately 176, as compared to 208 followers per Twitter user~\cite{Smith:2013}. With only one-tenth the number of users as Twitter, this average number of followers depicts that the Pinterest network is very-well connected.

\begin{table}[!h]
\centering
    \begin{tabular}{p{1.5cm}p{1.4cm}p{4.55cm}}
    \hline
{\bf Followers} & {\bf Followees} & {\bf Interests / Profession}                                 \\ \hline
11,992,745 & 149     & Designer / Blogger / Food                                     \\
9,099,998 & 143      & Designer / Magic / Food                                     \\
8,056,723 & 1,176    & Interior Designer                                    \\
7,519,854 & 205       & Not Mentioned                            \\
6,004,793 & 1,106     & Lifestyle Blog                                       \\
5,023,007 & 242       & Beauty Enthusiast / Blogger                            \\
4,793,914 & 310       & Architecture Student/Blogger                         \\
4,409,097 & 66         & Not Mentioned                                                   \\
4,126,895 & 1,001    & Artist                                               \\
3,658,844 & 383       & Freelancer / Blogger                                   \\ \hline
    \end{tabular}
     \caption{Top 10 user profiles on Pinterest based on number of followers (as of January, 2013). The table also shows number of followees for users, and interests / profession as captured from the about field.}
\label{tb:top_profiles}
\end{table}

\begin{figure*}[ht]
\centering
       	\subfigure[]{%
            \label{fig:follower}
            \includegraphics[scale=0.25]{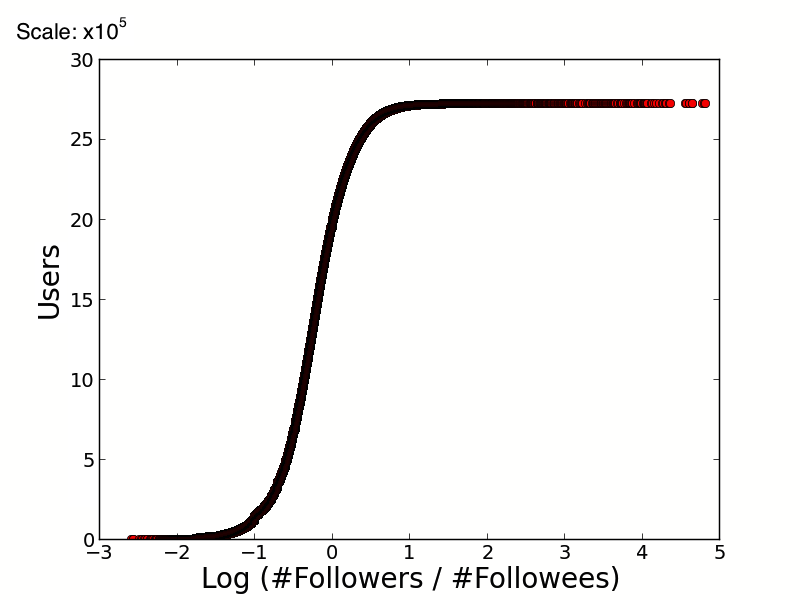}}
\subfigure[]{
    \label{fig:folpintw}
    \includegraphics[scale=0.25]{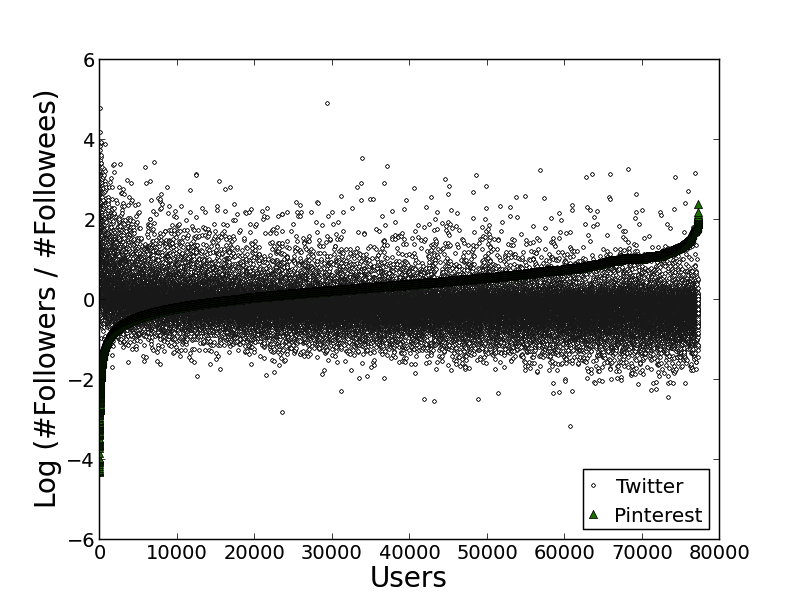}}
\subfigure[]{
    \label{fig:pinDesc}
    \includegraphics[scale=0.29]{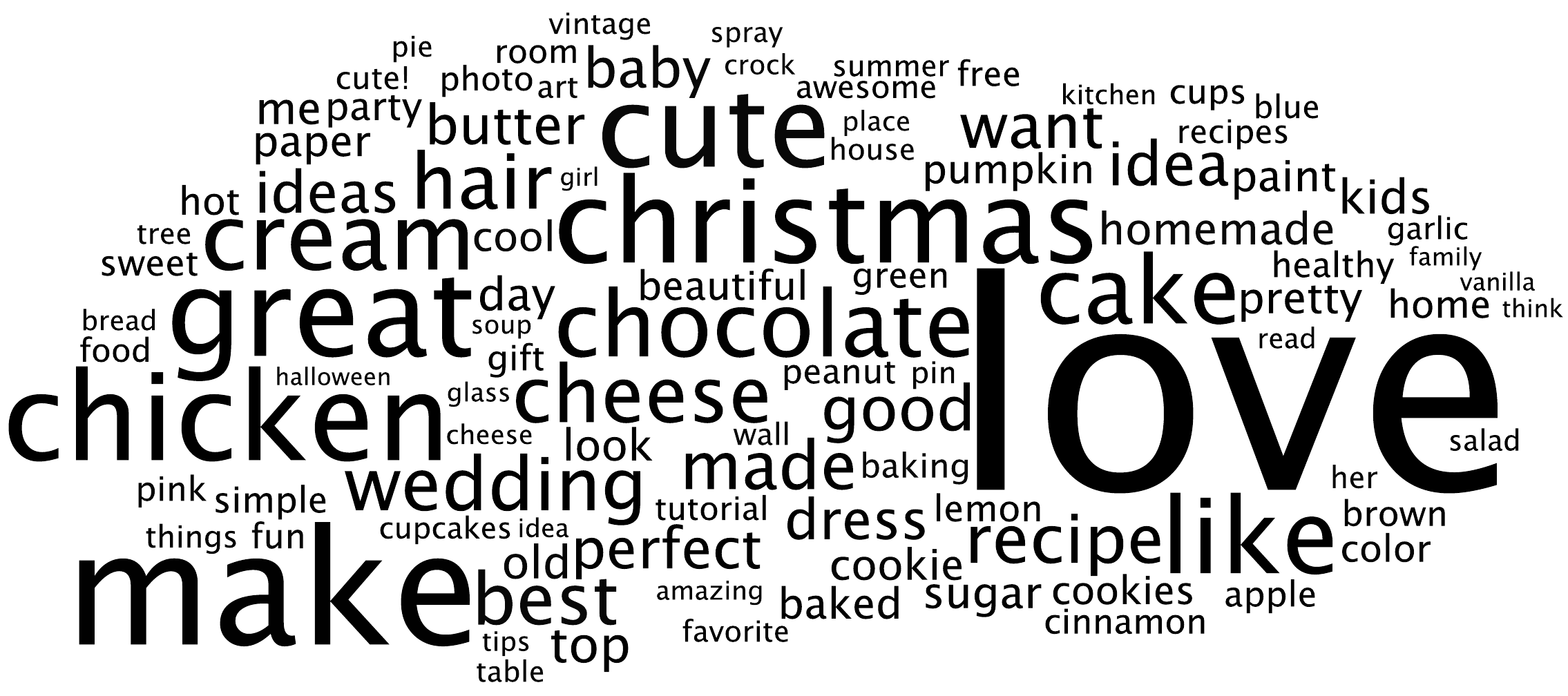}}

\vspace{-7pt}
\caption{(a) Followers / followees for the users on Pinterest, on a log scale. (b) The follower / followee ratio on Pinterest had no correlation with the ratio on Twitter. (c) Pin description on Pinterest. Similar to user descriptions, pin descriptions were also dominated by terms related to food and creative arts, and partially overlapped with terms present in user descriptions.}

\end{figure*}

We then plotted the ratio of number of followers versus the number of followees for all users (except for the users with 0 followees) on a log scale as shown in figure~\ref{fig:follower}, and found that more than 70\% users had more followees than followers. The graph depicts that a very small fraction of users had this ratio skewed, and most users on Pinterest in our dataset had a comparable number of followers and followees. Krishnamurthy et al.~\cite{krishnamurthy2008few} found a similar relation between followers and followees for Twitter users.

From the 328,570 users who had connected their Twitter accounts with Pinterest, we extracted the number of Twitter followers and followees for 93,659 users. We then plotted the ratio of followers / followees for these users for both, Pinterest and Twitter, on a log scale, as shown in figure~\ref{fig:folpintw}. As the plot suggests, the ratio of followers / followees on Pinterest was weakly correlated with the ratio of followers / followees on Twitter (correlation $=$ 0.32). Users who were popular on Pinterest were not necessarily popular on Twitter (and vice versa).

\subsubsection{Gender distribution} \label{sec:user:gender}
We extracted gender information from Facebook profiles of over 1.85 million users who had linked their Pinterest profiles with Facebook. Over 1.61 million users (87.15\%) were females, and only 130,945 users (7.04\%) were males. The rest (5.81\%) did not have their gender information publicly available. This gender distribution is quite similar to the one observed by Ottoni et al. in their work on Pinterest~\cite{ottoni2013ladies}.

\subsection{Pin characterization}

\subsubsection{Pin description} \label{sec:pindisc}
To understand the most common type of pins on Pinterest, we extracted the textual content present in the ``pin description" fields from all the pins, and analyzed the most frequently occurring terms. Figure~\ref{fig:pinDesc} represents the tag cloud of the top 100 terms present in pin description. Similar to user descriptions, terms related to food and creative arts dominated the pin description. We observed that 35\% of the words so obtained were directly related to food, 49\% of such food related words were found to be cooking ingredients, such as butter, oil, garlic, vinegar, spices etc. Here are some of the pin descriptions: \emph{``Recipe for Olive Garden Salad Dressing! I love this stuff!'', ``28-pc. Cupcake Bakeware Kit by Junior MasterChef'', ``Avocado Salsa! This stuff is incredible to top (or dip) your favorite Mexican food in!'', ``Video: pastry chef Joanne Chang takes us step by step through her foolproof recipe for French macarons'', ``15 Foods that Boost Your Metabolism!.''} This shows that Pinterest is heavily used to talk about different varieties of food, and to discuss about various recipes online.

Other than food, decoration and wedding related pins were also found to be very common. For example, \emph{``Printable Snowflake Wedding Invitations'', ``Silk Bride Bouquet Peony Flowers Pink Cream Lavender Shabby Chic Wedding Decor. \$94.99, via Etsy.'', ~``Wedding dresses and bridals gowns by David Tutera for Mon Cheri for every bride at an affordable price Wedding Dress Style'', ``Vintage Wedding Decorating Ideas''.}

\subsubsection{Statistics and topical analysis}
From our dataset of over 58 million pins, the average number of pins per user was 444.86 (min $=$ 0, max $=$ 100,135). The average number of repins per pin was found to be 0.72 (min $=$ 0, max $=$ 20,212). 
Almost 79\% pins in our dataset never got repinned. 
The average number of likes per pin was 0.21 (min $=$ 0, max $=$ 5,640). Also, 90.32\% pins were not ``liked" by anyone. This low percentage of repins and likes shows that there is a limited set of pins that get popular, and that a majority of pins go unnoticed. 
In case of comments, the results are even more skewed compared to pins. The average number of comments on a pin was 0.0065 (min $=$ 0, max $=$ 3,345), and 99.53\% pins had no comments. This shows lack of utility of the comment feature on Pinterest.
Comment feature on Pinterest allows the user to critique or remark on the already existing pins; this helps in establishing a connection between various users. A person need not follow the profile of the user in order to comment on his / her pin. Intuitively, this allows diverse and unbiased opinions that can be posted about a particular image. 

\begin{figure}[!h]
\centering
\includegraphics[scale=0.32]{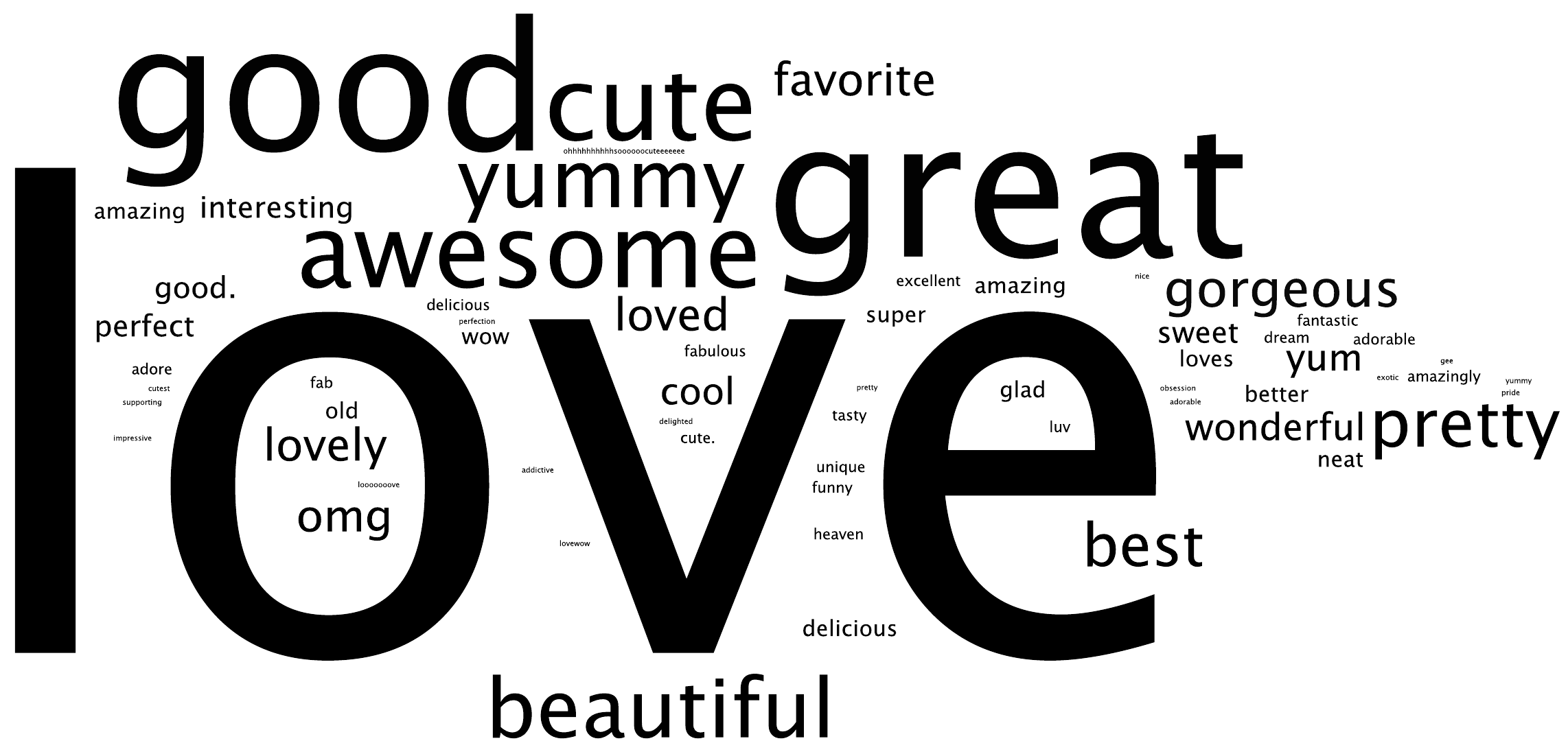}
\caption{Tag cloud of the top 63 positive words taken from comments on Pins.}
\label{fig:Comments}
\end{figure}

In order to study this diversity, we collected the comments for a total of 643,653 unique pins from our dataset. We then analyzed the kind of comments made by manually grouping them as ``positive comments'' and ``negative comments''. Here, we define positive comments as the ones which include some praiseworthy or optimistic remarks, and negative comments as the ones that show some kind of dislike or doubt of credibility for the pin. In order to filter these comments, we performed a basic string matching with some predefined set of words like awesome, nice, beautiful, wonderful, perfect, wow, and amazing etc. for positive comments; and bad, worst, hate, poor, dislike, and spam etc. for negative comments. Some positive comments were: ``Cutest puppy ever!!Sausage pants. Awesome.Love that sweet expression!'', ``Love this Amazing! Is it taken with your camera?? Made me smile:)yesss This is my kind of taste'', ``I love the perfect pastel colors of this bouquet.'', ``its looking yummy''. Figure \ref{fig:Comments} highlights the most frequently used positive words that we extracted from the comments. Some negative comments: ``Gross yuck yuck and yuck'', ``I hate people... I rrrreaaalllyy do.. :P'', ``spam''.
Through this analysis, we observed that only a small fraction of comments were found to be negative comments. 
Also, not all the comments incorporating the negative words defined above were actually negative. Some of the examples are: ``I want this so bad!  I think it is so precious!!!!!'', ``Bad ass!'', ``I want a goat do(so) bad!'', ``poor dog''. Comments that mention some particular pin as a spam highlights the existence of spam on Pinterest.

\begin{figure*}[!ht]
\centering
\includegraphics[scale=0.43]{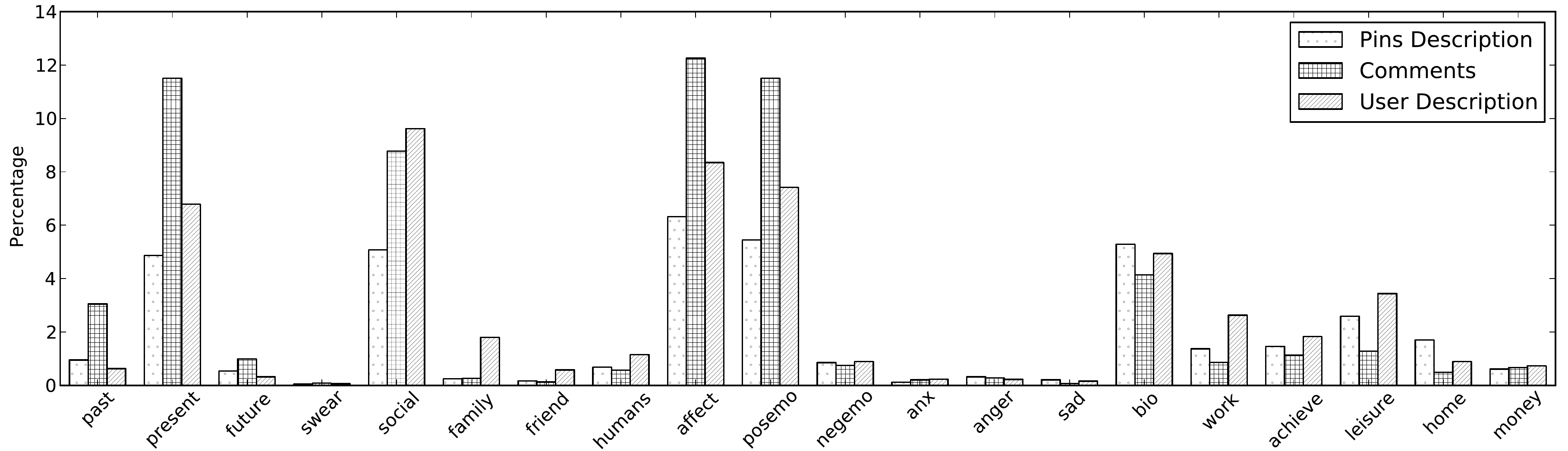}
\caption{LIWC analysis of textual content on Pinterest. Majority of the content comprised of positive sentiment words, or words indicating social interactions.}
\label{fig:liwc}
\end{figure*}

To get deeper insights about the content of these comments, we randomly crawled 643,653 (1.1\%) pins from our pin dataset, and were able to extract 2,544 comments. 
We then applied the Linguistic Inquiry and Word Count (LIWC) tool~\cite{pennebaker2007development} on these comments, pin descriptions (Section~\ref{sec:pindisc}), and user profile descriptions (Section~\ref{subsec:profiledescription}). We found that a large portion of the comments reflected positive emotion (Figure ~\ref{fig:liwc}). A similar pattern of positive emotion was observed for user description, as well as board names. In general, the network was found to have a large fraction of social content suggesting active human interaction. Presence of sad emotion, anger, anxiety, and swear words was found to be minimal. People discuss less about their past or future, and more about the present. Textual content depicting biological processes, work, and leisure activities was also found in substantial quantity. 
From all this analysis, we conclude that a user usually leaves a positive remark for a pin on Pinterest, and posts positive textual content in general.

\subsection{Source Analysis}\label{section:source}

Each pin has a \emph{source} embedded in it. This source is the ~original URL of the image from where it is ``pinned".~\footnote{Example of an image source:\url{www.cookingchanneltv.com/recipes/spanish-tortilla-recipe/index.html}} 
However, if the user has directly uploaded an image to Pinterest, the source field is set as ``pinterest.com". 
Table~\ref{tab:alexa} shows that the top source for images on Pinterest is the users themselves, i.e. a large portion of images are directly uploaded and pinned by the users. Out of all the pins in our dataset, 2,768,851 pins (4.7\%) were uploaded by users, second spot was taken by Google, which included images from Google Image Search, and other Google products, followed by Etsy, at the third spot. 
Not surprisingly, free image sharing platforms dominated the top 10 sources. 
Six out of the top 10 sources on Pinterest were among the top 1,000 most visited websites in the world~\cite{Alexa-Internet:2013}. Etsy, a commercial website being ranked high, shows that a reasonable amount of user traffic on Pinterest comes from e-commerce websites, and depicts that commercial activity is widespread on Pinterest.

Next, in order to explore secure pinning practices, we extracted the protocol information for these sources, and found that that 94.08\% of the sources used HTTP (Hyper Text Transfer Protocol), 1.2\% used HTTPS (Secure HTTP), and for 4.7\%, we could not get this information. Both the protocols, HTTP and HTTPS are used for transmitting and receiving information across the Internet, but HTTPS is more secured and provides an encrypted connection to the server~\footnote{\url{http://theprofessionalspoint.blogspot.in/2012/04/http-vs-https-similarities-and.html}}. Therefore, from the above statistics we infer that majority of the pin sources used on Pinterest were unsercured.

\begin{table}[!ht]
\begin{centering}
    \begin{tabular}{llll}
    \hline
Source        & Count    & W.A.R.                
& Category
\\ \hline
Pinterest.com          & 2,768,851 & N/A                  
& N/A
\\
Google        & 1,293,749 & 1                    
& Search engine
\\
Etsy          & 1,157,815 & 164                  
& Commercial
\\
Flickr        & 625,686   & 70                   
& Image sharing
\\
Tumblr        & 486,984   & 31                   
& Image sharing
\\
Imgfave       & 376,179   & 9,462                
  & Image sharing
\\
Weheartit     & 306,443   & 970                  
& Image sharing
\\
Someecards     & 296,908   & 6,648             
 & E-cards
\\
Houzz         & 294,065   & 958                  
& Home decor.
\\
Marthastewart & 292,128   & 2,439            
    & Food / Art
\\
\hline
    \end{tabular}
    \caption {Top 10 image sources on Pinterest. W.A.R.= Worldwide Alexa Rank. Apart from free image sharing / social network platforms, top sources include commercial platforms like Etsy.}
    \label{tab:alexa}
\end{centering}
\vspace{-5pt}
\end{table}

%


\subsection{Pinboard analysis}\label{section:pin_board}
\begin{figure*}[!ht]
\centering
\includegraphics[scale=0.45]{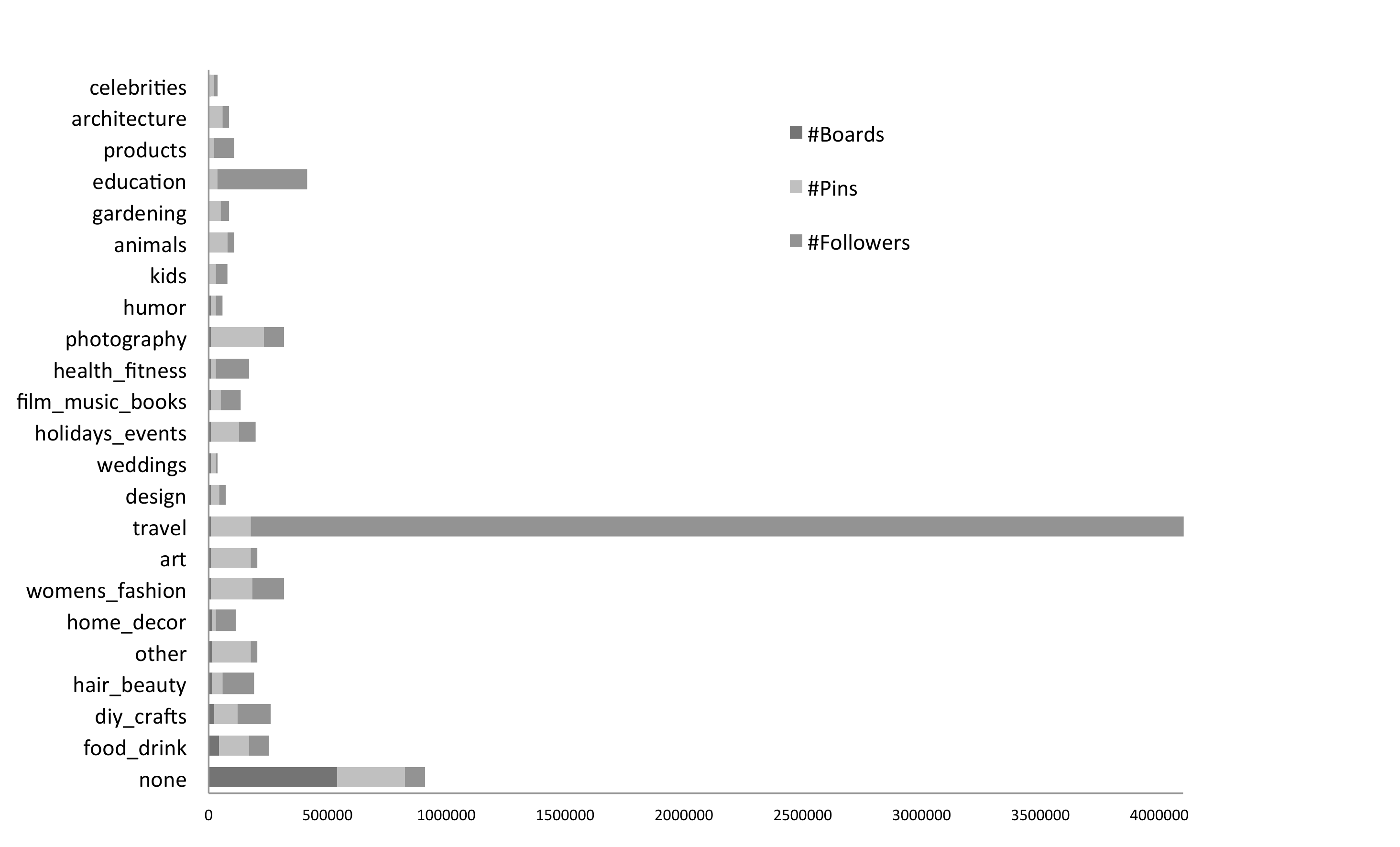}
\caption{Board categories on pinterest. Y axis represents the predefined board categories and X axis represents the number of boards, pins, and followers under each of these categories.}
\label{fig:category}
\end{figure*}

In addition to the above Pin analysis, we also analyzed the names of Pinboards. 
The most common terms occurring in board names were home, style, recipes, food, wedding, crafts, etc. Pinterest also provides an option with 33 different predefined categories for board creation. We analyzed the popularity of all these categories based on 3 factors, \emph{number of boards in each category, number of pins on these boards}, and \emph{number of followers of these boards} under each category. We saw that 69.37\% boards were created with no standard category selected. Apart from these, the top three categories for board creation were food\_drink (5.6\%) ~followed ~by diy\_crafts (2.3\%), and hair\_beauty (2.4\%). Followers of boards in the ``travel" category outnumbered all the other boards by a big margin (Figure~\ref{fig:category}), and had the highest ratio of followers per pin (23.69 followers per pin). The next most famous boards in terms of followers per pin were education (10.34 followers per pin), health\_fitness (5.37 followers per pin), and home\_decor (4.71 followers per pin). 



\subsection{Location analysis}
\begin{figure}[!ht]
\centering
\fbox{\includegraphics[scale=0.3]{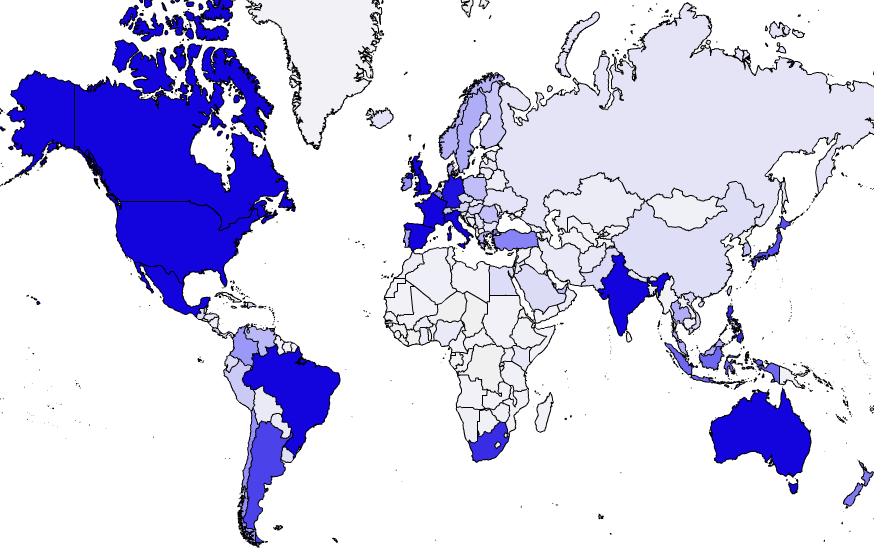}}
\caption{World location heatmap on Pinterest. Darker regions indicate the presence of more Pinterest users. Very little traffic comes from Africa (except South Africa), Russia, and China.}
\label{fig:heat}
\end{figure}

We investigated location information to find the Pinterest population distribution across the world. 
From our dataset, we collected 192,261 valid user locations, and performed a lookup using Yahoo PlaceFinder API. We inferred the top 10 countries in terms of number of users (Table~\ref{tb:top10}) from Yahoo's API output. Similar to Facebook and Twitter~\cite{krishnamurthy2008few}, a majority of Pinterest users also came from the U.S.A., Canada, U.K., Brazil, India, and Europe; Figure~\ref{fig:heat} shows the corresponding heatmap. We found minimal users from Africa, Russia, and China.
Table~\ref{tb:top10} also lists Pinterest's regional traffic ranks taken from Alexa, on 2nd June 2013. These ranks show that Pinterest is among the top most popular sites in countries like U.S.A., Canada, U.K., Australia, Brazil, India, etc., which are also the top user locations in our dataset. After analyzing country-wise distribution, we did a city level location analysis for these top 10 countries (Table~\ref{tb:top10}), and found that most Pinterest users belonged to big metropolitan cities. More than half of the cities in top 20 were from the U.S.A. Pinterest's penetration was found to be quite low in smaller cities. 

\begin{table*}
\centering
    \begin{tabular}{llccc||llc|llc}

    \hline
    \multicolumn{5}{c||}{{\bf Countries}} & \multicolumn{6}{c}{{\bf Cities}} \\
    \hline
  \multicolumn{2}{c}{Country}     & P.R.R. & Females (\%) & Males (\%) & \multicolumn{2}{c}{World City}  & Count & \multicolumn{2}{c}{World City} & Count \\ \hline

1. &    U.S.A       & 15     & 83.88      & 8.80     & 1. & New York      & 5597  &  11. & Dallas       & 1275  \\

2. &    Canada      & 21     & 82.73      & 10.66    &  2. & London        & 3424  &  12. & Austin       & 1249  \\

3. &    U.K.        & 38     & 72.79      & 18.47    &  3. & Los Angeles   & 3194  &  13. & San Diego    & 1213  \\

4. &    Australia   & 23     & 80.59      & 11.05    &  4. & Chicago       & 2593  & 14. &  Houston      & 1169  \\

5. &    Brazil      & 73     & 73.94      & 18.47    &  5. & Toronto       & 1752  &  15. & Sidney       & 1157  \\

6. &    Spain       & 54     & 66.83      & 24.56    & 6. &  San Francisco & 1659  & 16. &  Paris        & 1078  \\

7. &    Italy       & 142    & 62.91      & 27.04    &  7. & Atlanta       & 1472  & 17. &  Melbourne    & 1034  \\

8. &    France      & 183    & 70.36      & 22.53    & 8. &  Washington    & 1428  &  18. & Portland     & 1010  \\

9. &    India       & 20     & 45.30      & 46.64    &  9. & Seattle       & 1332  &  19. & Vancouver    & 959   \\

10. &    Netherlands & 29     & 75.88      & 16.52    &  10. & Boston        & 1329  &  20. & Philadelphia & 851   \\ \hline

    \end{tabular}
\caption{Top 10 countries, and top 20 cities in decreasing order of Pinterest population. Apart from India, all other countries were dominated by female users. The penetration of Pinterest is maximum in big metropolitan cities. P.R.R.= Pinterest Regional Rank.}
\label{tb:top10}
\end{table*}
%
%

As most Pinterest users in our dataset were females (Section~\ref{sec:user:gender}), we analyzed gender distribution with respect to location. We observed that approximately 88\% of users from the U.S.A. were females, and approximately 7\% were males. A similar trend was observed in U.K., Australia, Europe, and Brazil (Table~\ref{tb:top10}). India was the only country in the top 10, where the number of male users (46.64\%) was greater than the number of female users (45.30\%). 



\section{Privacy and Security}\label{section:privacy_security}
In this section, we present our security and privacy analysis on Pinterest.
\subsection{PII leakage}
Trust and privacy have been previously studied as a concern on online social networks~\cite{dwyer2007trust}.
Pinterest is no different when it comes to such user privacy concerns online. The feature for linking Facebook and Twitter accounts with a Pinterest account has been reported to breach privacy~\cite{Lupfer:2012}. Moreover, since all user information and activity on Pinterest are completely public, the risk of leakage of privacy is high.
Recently, users of Pinterest requested the site to implement a feature where they could keep some of their content private~\cite{Blog:2012}. In response to these requests, Pinterest introduced ``Secret Boards" (introduced in November 2012), whose content (the board and it's pins) are visible only to the users who create them.

Users' request for Secret Boards highlight the need for features to better control users' privacy on Pinterest. However, apart from these secret boards, all other user data collected by Pinterest like profile picture, profile description, location, Facebook and Twitter ID are publicly available. In our dataset, we found many instances where users gave out private information about themselves in the profile description field. 
For example, a user in her profile description wrote: \emph{``I am 20 years old and I am going to school to become a nurse! I am getting married May 25 2013 and can't wait!"}. Another user stated, \emph{``Mommy of two cute little kids and wife to a marketing man living in Portland, Oregon."} 
Research shows that it is possible for third-parties to link  PII, which is leaked via OSNs, with user actions both within OSN sites and elsewhere resulting in privacy leakage~\cite{krishnamurthy2009leakage}.

We found that a total of 9,926 users in our dataset shared their email addresses publicly.  
We then searched for phone numbers, which are widely considered to be PII~\cite{kumaraguru2012privacy}, and found a total of 1,046 phone numbers and / or BBM pins from the users' profile description field. For example, a user wrote \emph{``I am a Doctor in Unani medicine (BUMS) -- Medical Officer at Govt. Unani Dispensary, Manjeri. Chief Consultant at KERALA [A state in southern part of India] UNANI HOSPITAL ~Opp ~KMH, ~Manjeri ~Ph:+91~ 938XX 2XXXX ~www.keralaunani.com".}

\subsection{Copyright Infringement on Pinterest}\label{section:copyright}

Copyright is a form of protection provided to authors of ``original works of authorship, including pictorial, graphic, and sculptural works"~\cite{Office:2009}. 
Copyright infringement occurs when a copyrighted work is reproduced, distributed, performed, publicly displayed, or made into a derivative work without the permission of the copyright owner. Application of copyright laws holds true for all artistic work shared on Pinterest. When a user pins an image (from another website or a manual upload), the image gets copied to the Pinterest server. Such an image may or may not be in public domain and thus, can invoke creators to lodge a complaint under Copyright Acts (e.g. U.S. Copyright Act~\cite{Office:2009}). Many users give due credit or link back to the source, but an infringement may arise when users don't give due credit to the creator of the image / pin~\cite{case1,case2} (we use the terms ``pin" and ``image" interchangeably). In this section, we explore the possibilities of such cases of copyright infringement by extracting and analyzing copyright and authorship data from images. 
In order to extract this information, we performed analysis on the Exchangeable Image File Format (EXIF) data from a random sample of 9,950 images. We found copyright information in 1,415 images, and artist information in 1,587 images. We then picked a random sample of 100 images containing artist information, for further analysis. 

With information extracted from EXIF data, it was only possible to check if the copyright holder had herself shared the image online, and if other people sharing the image, had given her due credit. A user can credit a pin by either mentioning the name of copyright holder in the pin description, or citing the original source URL. 
In order to determine if the copyright holders get credited on Pinterest or not, we manually analyzed 100 random pins based on 5 parameters: (1) whether the pin has been pinned from some website or uploaded by the user; (2) did the pinner give any credit to the copyright holder of that pin in its description; (3) if the original pinner is the copyright holder of that pin; (4) did the source website credit the copyright holder; and (5) if the source website belonged to the copyright holder. 

\begin{figure}[!h]
\begin{centering}
\fbox{\includegraphics[scale=0.4]{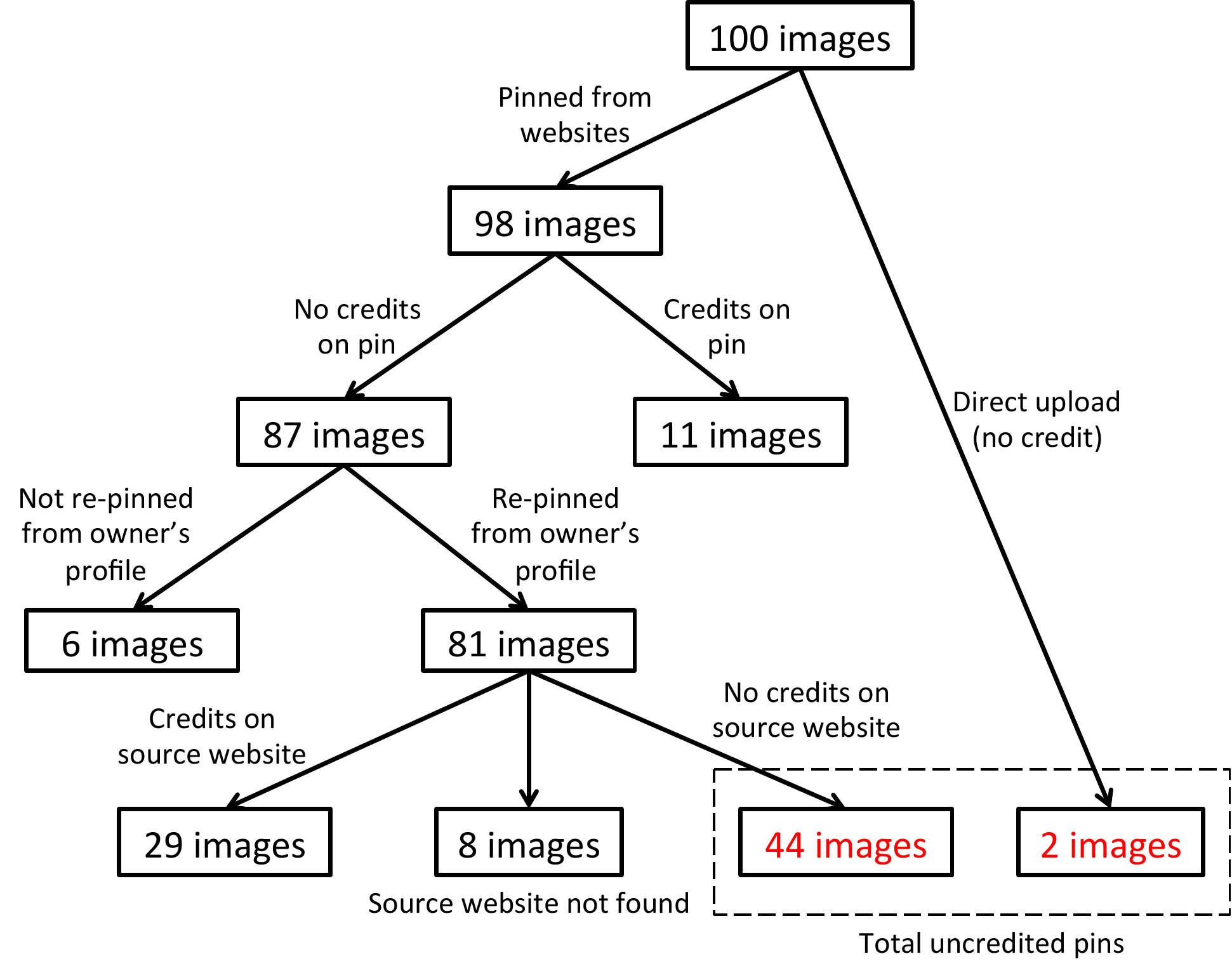}}
\caption{Decision tree diagram to check for credited images on Pinterest. We found 46 out of 100 random pins uncredited.}
\label{fig:100copy}
\end{centering}
\end{figure}

Figure~\ref{fig:100copy} represents the decision tree diagram for the methodology we followed for our manual verification process. We observed that out of 100 images that we analyzed, 98 were pinned from various websites and only 2 pins were directly uploaded. These directly uploaded pins did not have any credits to the copyright holder in their pin description. Of the 98 other pins, only 11 gave credits in the pin description. For the pins which did not mention credits (87), we found that 6 were originally pinned from the Pinterest profile of their copyright holder. To verify this, we simply compared the name of that Pinterest profile user with the name of the copyright holder from our EXIF data (though we don't claim this information to be 100\% true, since names are not unique). We considered these pins to be credited, and were finally left with 81 pins which were neither pinned from the profile of the copyright holder, nor the holders being given credit in the pin descriptions. Analyzing their source websites, we observed that 44 (54.32\%) of these 81 pins did not contain any credits on their source websites as well (for 8 of these 81 pins, source websites no longer exists, so we could not check for them). Thus, a random sampling of 100 pins gave a total of 46 (44 with source websites, and 2 directly uploaded) non-credited pins. This highlights that a large proportion of digital work that exists on Pinterest goes uncredited; this can be seen as one of the problems leading to copyright issues on Pinterest. 
We believe the 5 parameters we selected, cover all methods of checking whether a pin is credited or not. However, our methodology to detect copyright infringement depends only on manual verification and confirmation based on the EXIF data. 

\begin{figure}
\centering
\includegraphics[scale=0.7]{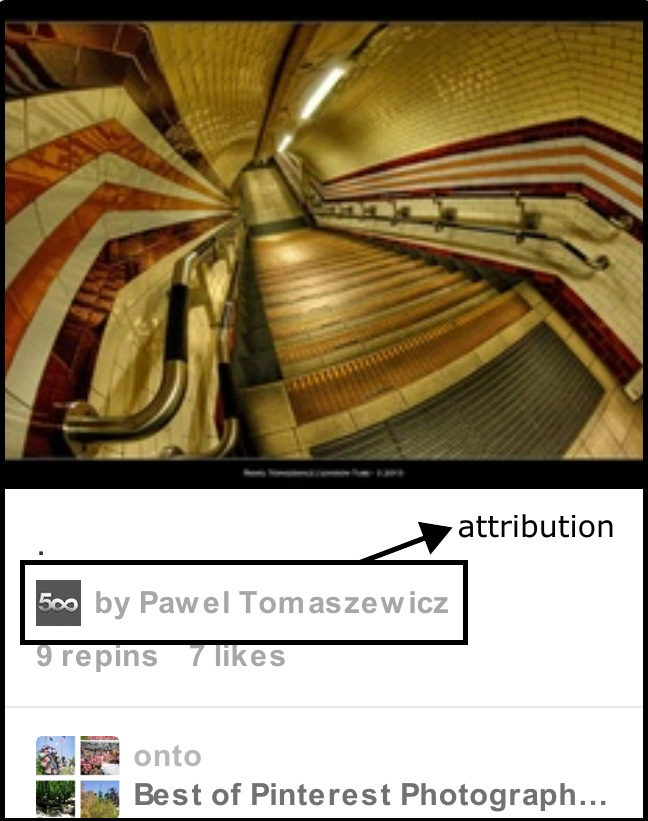}
\caption{Attribution feature on Pinterest.}
\label{fig:attribution}
\end{figure}

Pinterest also provides a feature called \emph{attribution} (Figure ~\ref{fig:attribution}). With this feature, an attribution statement gets added to a pin when a user pins from some specific sources like Flickr, YouTube, Vimeo, Behance, 500px, Etsy, Kickstarter, SlideShare, and SoundCloud. Such attribution statements (consisting of author name and the original source website) can neither be removed, nor edited. These attributions are placed by Pinterest to ensure due credit is given to the copyright holder and can be used by any website that hosts original content.~\footnote{\url{https://help.pinterest.com/entries/21356541-Crediting-pins-on-Pinterest}} 
If a user discovers a pin from some website where the content is not originally uploaded, Pinterest tracks back and determines the original source (if the source is attributed). This is known as deep linking~\cite{wassom1998copyright}. Although such copyright protection measures already exist, a very few websites are actually using it. This is also clear from our aforementioned 100 pin analysis, where almost half of the pins were found to be uncredited. Such measures like attribution and deep linking, if adopted widely, can reduce the existing copyright infringement problems, 
and can ensure the originators of these digital arts get the due credit for their creative work.

\subsection{Malware Analysis}
While various brands are using Pinterest for legitimate commercial purposes by promoting their work through pinboards, Pinterest has also attracted spammers and malicious users. With the growth in the number of users, there has been a simultaneous growth in the number of spammers on Pinterest.~\footnote{\url{http://mashable.com/2012/12/06/pinterest-spam-accounts/}} Numerous online scams have been reported in recent times~\cite{Cluley:2012,Alerts:2012,Jelea:2012,Pichel:2012}, and Pinterest has taken measures to solve this problem.~\footnote{\url{http://blog.pinterest.com/post/37347668045/fighting-spam}}
To get a better understanding of the presence of spam and malware on Pinterest, we used \emph{Google's Safe Browsing API}~\footnote{\url{https://developers.google.com/safe-browsing/}} to check for malicious source URLs on the network. We analyzed the source URLs of a random sample of 5.5 million pins from our pin dataset and found 1,322 (0.024\%) unique malware pins. Despite numerous reported incidents of spam and malware, such low number suggests that the techniques deployed by Pinterest to avert malware are indeed effective. 
Since we collected these pins in January 2013, we wanted to check if the captured malware continued to exist on Pinterest. We then crawled these 1,322 pins again in May 2013 
and observed that 33 of these pins no longer existed. It is hard to predict if the users themselves deleted these pins, or Pinterest removed it. 
We re-checked the source URLs of the 1,322 malware pins in May 2013, and found that 223 source  URLs no longer exist. Corresponding to the 1,322 malware pins, we identified 1,171 unique users from our dataset. Re-crawling these user accounts in May 2013 revealed that 100 out of these 1,171 user accounts did not exist. This shows that other than removing malicious content, Pinterest also take measures to remove malicious user profiles. 
To find the most infected category of boards, we determined the categories of boards on which majority of such malware pins were posted. Table~\ref{tb:malware_boards} shows the top 10 board categories, with the number of malware pins associated with each of them. Interestingly, this distribution does not follow the general popularity distribution of board categories on Pinterest. The most popular board categories found in Section~\ref{section:pin_board} were travel, education, and health\_fitness which were not the most infected boards; categories like My Style, Favorite Recipe, Food and Drink were found to have the maximum number of malware pins. One possible explanation for this pattern could be a targeted attack, where the intent of users spreading malicious content is not to target the most popular boards, but to target a specific set of users or boards.
Similar to legitimate pins, malware pins were also found to have very few likes, repins, and comments. From the 1,322 malware pins, only 2 pins had more than 10 likes (31 and 16); over 90\% (1,202) pins had no likes. Maximum number of repins were found to be 60 for a pin, and 1,019 (77.08\%) pins had no repins. In case of comments, only 1 pin had 6 comments, and 1,316 (99.54\%) pins had no comments. These observations suggest that both the presence and diffusion of malware on Pinterest is very limited. Only a small number of users are affected by / related to malware on Pinterest.

\begin{table}[!h]
\centering
    \begin{tabular}{lc}
\hline
 {\bf Board Name}                 & {\bf Malware Pins} \\
\hline
My Style                   & 95            \\
Favorite Recipe            & 76            \\
Food and Drink             & 56            \\
Craft Ideas                & 42            \\
For the Home               & 36            \\
Wedding Ideas              & 30            \\
Books Worth Reading        & 21            \\
Favorite Places and Spaces & 14            \\
Christmas                  & 14            \\
Drinks                     & 14            \\
\hline\\
\vspace{-20pt}
    \end{tabular}
  \caption{Top 10 board categories in terms of number of malware pins. These 10 board categories contributed to 30.1\% of the total malware pins in our dataset.}\label{tb:malware_boards}

\end{table}

Table~\ref{tb:malware_source} lists the top 10 sources of malware pins, along with the corresponding number of pins from these sources. Top 10 sources contributed  821 (62.1\%) out of the 1,322 malware pins. 
Revisiting these sources in May 2013 revealed that apart from one out of the 10 sources, all other sources continued to exist. Table~\ref{tb:malware_source} also shows the number of pins from these particular sources that were removed as of May 2013. We observed that out of 821 pins from these top 10 sources, only 17 pins had been removed. This indicates that while Pinterest may be taking measures to counter malware on the network, it is not doing so very frequently.

\begin{table}[!ht]
  \centering
\begin{tabular}{lccc}
  \hline
{\bf Source } & {\bf SF} & {\bf MP }& {\bf PR } \\
  \hline
lilyboutique.com & Yes & 180 & 5 \\
somewhatsimple.com  & Yes & 103 & 1 \\
drinkeattravel.com & Blocked & 100 & 0\\
stilettostolegos.com & Yes & 87 & 1\\
cookiesandcups.blogspot.com & Yes & 79 & 2\\
safetydangerdefender.com & Yes & 75 & 4\\
savedbylovecreations.com & Yes & 63 & 1\\
beyondthebaby.com.au & No & 52 & 0\\
chocolatebrownierecipe.org  & Yes & 45 & 0\\
cookiesandcups.com & Yes & 37 & 3 \\
\hline
~ & {\bf Total} & 821 & 17 \\
  \hline
\end{tabular}
\vspace{-8pt}
  \caption{The top 10 sources of malware on Pinterest. Only 1 out of the top 10 source URLs was removed / blocked. SF = Source Found, MP = Malware Pins, PR = Pins Removed.}
\label{tb:malware_source}
\vspace{-9pt}
\end{table}



\section{Discussion}\label{section:discussion}

In this work, we characterized the Pinterest social network, and studied various aspects of security and privacy related to it. We collected 17,964,574 unique user handles, 3,323,054 complete user profiles, 777,748 boards with their corresponding details, and 58,896,156 unique pins with their related information, using Snowball sampling~\cite{goodman1961snowball}. Our analysis was based on a partial subgraph of the Pinterest network, and suggests that Pinterest is a social network dominated by ``fancy" topics like fashion, design, food, travel, love etc. across users, boards, and pins. A large part of the network was found to have a comparable number of followers and followees. Only a small fraction of people had large number of followers as compared to followees and vice-versa. The largest contributors of content (images) on Pinterest were the users themselves, with 2,768,851 (4.7\%) users uploading original content; the remaining content (95.3\%) was pinned from pre-existing web sources. Google Images, and Etsy followed as the next most famous sources, from where images are pinned onto Pinterest. USA, Canada, and UK contributed the maximum proportion of users, together accounting for over 73\% of the total Pinterest population. 

We then focused our analysis on the security and privacy issues on Pinterest, and found numerous instances of PII leakage through users' description field. Although this information is shared voluntarily by users, the \emph{all public} nature of the network makes users potentially more vulnerable to third-parties extracting and using this information for marketing and other purposes. We also performed analysis on EXIF data from pinned images, and discovered multiple potential cases of copyright infringement on Pinterest, where users failed to give credit to the owner of the content they shared. Although Pinterest provides features like attribution, only a small number of websites were making use of it to protect their content from copyright violations. Upon analyzing pins for malicious content, we found a small fraction of malware pins using the Google SafeBrowsing API lookup. We noticed that the boards with the most number of malware pins were not the most famous boards on Pinterest.


We picked the initial seeds for our data collection process as the top 5 most followed users on Pinterest. We understand that this technique suffers from community bias, and the sample taken is not completely random. To dampen the effect of this bias, we crawled only partial sub-graphs for all the 5 seed users. 
Similarly, on the next level of our BFS crawl, we did not crawl more than 48 followers for any user. Since there is not much prior work on Pinterest, we do not have enough academic literature to claim that our dataset is representative of the whole Pinterest population. However, the previous work by Ottoni et al.~\cite{ottoni2013ladies}, and a report from Engauge, a digital marketing agency~\cite{Engauge:2012}, show similar gender distributions for users, and similar topic distributions for boards and pins as our dataset.

Our analysis on Copyright violation was highly manual, and is based on the information available publicly, or using EXIF analysis. With the available data, it was difficult to actually look for Copyright infringement, thus we restricted our analysis only to check for the presence of credits on pins. We also consulted some law experts in India for understanding the legal aspects of copyrights and their infringement. We would like to do a more comprehensive, and expert analysis in future, with a larger dataset. Another implication of our results lies in the methodology we used for malware analysis. These results are based purely on the source URLs of the images posted on Pinterest, and their status on Google's SafeBrowsing Lookup API. Given that Pinterest is an image-based social network, presence of image spam is a likely phenomenon. However, detection of image spam is an area of research in itself, and is out of scope of this work. An important feature present on Pinterest is the ability to view all pins from a particular source. For example, if a user who is logged into her account on Pinterest visits the URL \url{http://pinterest.com /source/flickr.com/}, the user gets to see all pins which have their source as \url{flickr.com}. We would like to further explore this feature to create security systems to monitor incoming pins from a particular malicious source and help identify suspected spammers, and phishers on Pinterest.



To the best of our knowledge, this is the first attempt to characterize Pinterest, and study its various components in depth, on such a large scale. However, we only present analysis for a partial dataset in this paper. For this analysis, we use profile information for only about 3.3 million users from over 17 million unique user handles that we had in our dataset. Our data collection process is still active, and we would like to redo our analysis on the largest connected component (LCC) of the Pinterest network. We would also like to perform a more detailed analysis of image-spam, and copyright violations on this network. 
Given that Pinterest has been the fastest growing social network in recent times, it would be interesting to see if malicious users are targeting Pinterest for spiteful purposes.

%

\section{Acknowledgements}
We would like to express our sincere thanks to all members of Precog research group at IIIT-Delhi, for their continued support and feedback on the project.
\bibliographystyle{abbrv}	
\bibliography{pinterest_bib}
\end{document}